\documentclass[aps,prl,twocolumn]{revtex4}  
\usepackage{amsmath}
\usepackage[pdftex]{color, graphicx}

\begin{document}
\title{Controlling the second-harmonic in a phase matched negative-index metamaterial}
\author{Alec Rose, Da Huang, and David R. Smith}
\affiliation{Center for Metamaterials and Integrated Plasmonics, Duke University, Durham, North Carolina 27708, USA}

\begin{abstract}
Nonlinear metamaterials (NLMMs) have been predicted to support new and exciting domains in the manipulation of light, including novel phase matching schemes for wave mixing. Most notable is the so-called nonlinear-optical mirror, in which a nonlinear negative-index medium emits the generated frequency towards the source of the pump. For the first time, we experimentally demonstrate the nonlinear-optical mirror effect in a bulk negative-index NLMM, along with two other novel phase matching configurations, utilizing periodic poling to switch between the three phase matching domains.
\end{abstract}


\maketitle

Nonlinear metamaterials (NLMMs) represent the pinnacle of designer electromagnetic materials, allowing one to manipulate light based on direction, polarization, frequency, and intensity. In particular, these artificial mediums have the potential to make nonlinear-optical phenomena accessible at low powers and in novel and exciting configurations~\cite{Pendry1999Magnetism, Zharov2003Nonlinear}. While a wide range of nonlinear processes have already been demonstrated in NLMMs, including frequency generation~\cite{Klein2006Second-harmonic, Shadrivov2008Tunable, Huang2011Wave}, parametric amplification~\cite{Popov2006Negative-index, Poutrina2010Parametric}, and bistability~\cite{Shadrivov2008Nonlinear, Wang2008Nonlinear}, the setups have been constrained to sub-wavelength interaction lengths for fear of the destructive effects of phase mismatch. As such, the conversion efficiencies achieved are only a fraction of their potential. However, phase matching is an exciting subject that involves the entire set of linear and nonlinear properties, bringing to bear the full capabilities of NLMMs to manipulate light at will. This claim has been supported by several theoretical studies on NLMMs, most notably the prediction of a nonlinear-optical mirror that generates second-harmonic (SH) \textsl{back} towards the fundamental frequency (FF) source~\cite{Agranovich2004Linear, Shadrivov2006Second-harmonic, Popov2006Negative-index}, but experimental verification of these novel phase matching effects in a bulk NLMM has been lacking.

In this letter, we experimentally demonstrate three unique phase matching configurations for second-harmonic generation (SHG) in a metallic waveguide loaded with a magnetic NLMM at microwave frequencies. The three configurations are reflected SH phase matching in a negative-index band, transmitted SH quasiphase matching (QPM), and simultaneous QPM of both the reflected and transmitted SHs near a zero-index band. The resulting SH spectrums, showing strong phase matching-induced enhancements, are supported qualitatively by nonlinear transfer matrix method (TMM) calculations on a homogeneous slab with equivalent effective properties.

\begin{figure}[!htbp]
\center
\includegraphics[width=.37\textwidth]{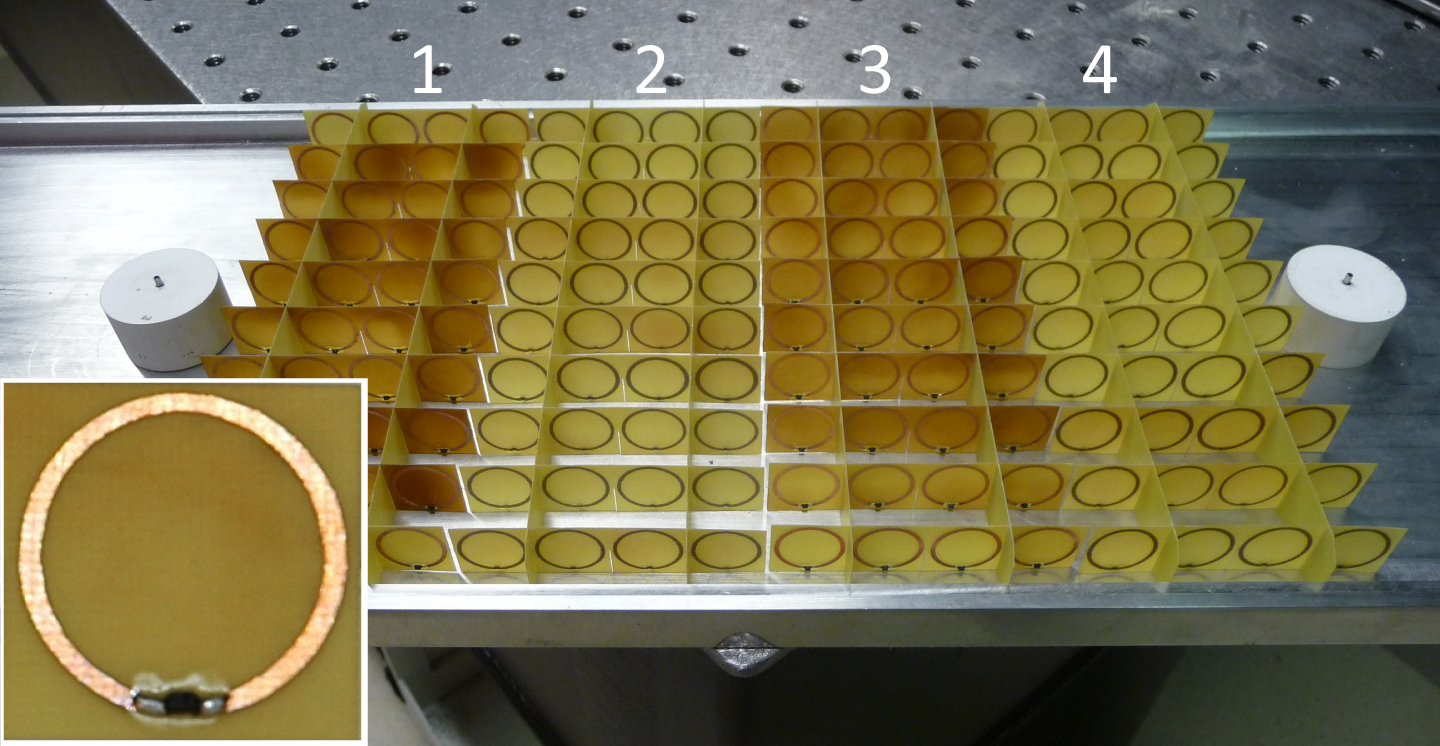}
\caption{(Color online) Photograph of the waveguide with the top removed, loaded with four identical sections of VLSRRs. The inset shows an enlarged view of the NLMM unit-cell.}
\label{waveguide}
\end{figure}

The NLMM under investigation in this letter is the varactor-loaded split-ring resonator (VLSRR), the canonical subject of numerous recent studies. The constituent unit-cell of this NLMM is composed of a 17 $\mu$m thick, 0.75 mm wide copper ring printed on a 0.2 mm FR4 PCB substrate. The 1 mm gap is loaded with a Skyworks SMV1231 varactor diode, whose capacitance varies with applied voltage according to $C(V_D) = C_0(1 - V_D/V_p)^M,$ where $V_D$ is the voltage across the diode, $C_0=2.35$ pF is the zero-voltage capacitance, $V_p=1.5$ V is the intrinsic potential, and $M=0.8$ is the gradient coefficient~\cite{Skyworks}. Our bulk material, displayed in Fig. \ref{waveguide}, consists of 16 layers (24 cm in the direction of propagation) of VLSRRs placed in an aluminum waveguide with cross-section 15 cm $\times$ 1.5 cm, completely filling the waveguide in the transverse direction, and arranged in a square lattice with 1.5 cm spacings. For the purpose of periodic poling, the VLSRR sample is divided into four identical sections that can be reoriented independently. The particular choice of dimensions ensures that the resonance frequency of the VLSRR is below the cutoff frequency of the waveguide's lowest-order mode. As such, the NLMM loaded waveguide is expected to support a backward mode at frequencies where the VLSRR's permeability is negative, that is, a mode whose phase velocity is opposite to the direction of power flow~\cite{Veselago1968Electrodynamics, Smith2000Composite, Marques2002Left-handed-media}. In addition, near the cutoff frequency, the wavevector of this mode vanishes, yielding a second region of interest for phase matching~\cite{Somerville2011Second}.

\begin{figure}[!htbp]
\center
\includegraphics[width=.43\textwidth]{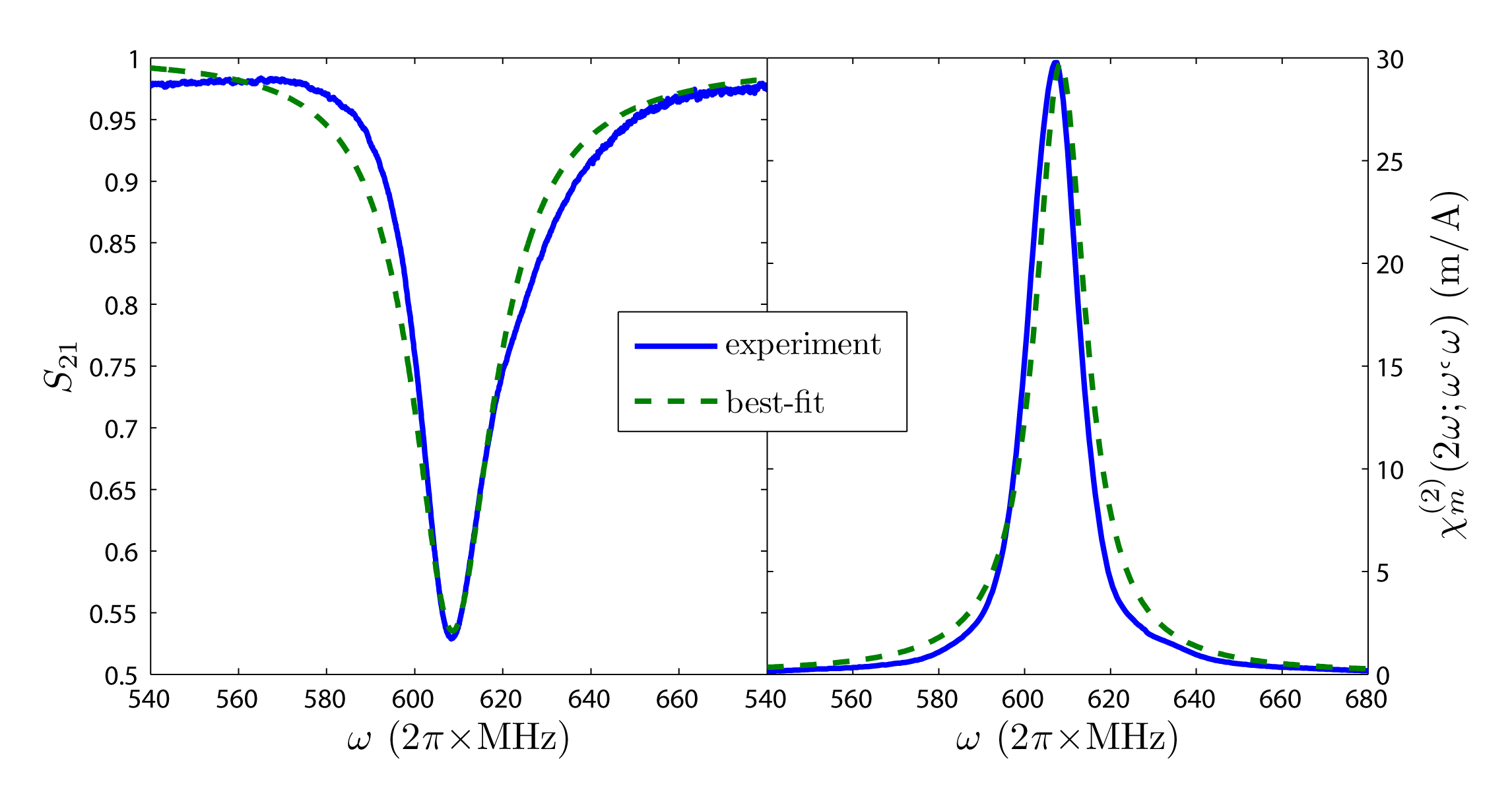}
\caption{(Color online) Plot of the magnitude of the retrieved (solid blue) and fitted (dashed green) $S$-parameters (a) and second-order susceptibility (b) for the VLSRR medium.}
\label{nl_retrieval}
\end{figure}

The linear properties of the VLSRR medium were characterized in a transmission line setup by measuring the transmission of a single layer and fitting to known property models~\cite{Larouche2010Experimental}, shown in Fig. \ref{nl_retrieval} (a). We find our VLSRR medium to be well-described by a constant permittivity $\epsilon_y(\omega) = 2.2 \epsilon_0$, and a permeability given by the Lorentz oscillator-like formula
\begin{equation}
\mu_x(\omega) = \mu_0\left(1 + \frac{F\omega^2}{\omega_0^2 - \text{i}\gamma\omega - \omega^2}\right), \label{mu}
\end{equation}
where $F=0.22$ is the oscillator strength, $\omega_0=2\pi \times 608$ MHz is the resonance frequency, and $\gamma=2\pi \times 14$ MHz is the damping coefficient. The permeability is assumed to be that of free-space along the other axes.

Subsequently, we measured the SHG output from this setup. Using the nonlinear TMM retrieval method~\cite{Larouche2010retrieval, Rose2010Nonlinear, Larouche2010Experimental}, we extracted the second-order magnetic nonlinear susceptibility from the SH spectrum (shown in Fig. \ref{nl_retrieval} (b)). The result was then fitted to an effective medium model obtained via a perturbative circuit analysis,~\cite{Poutrina2010Analysis}
\begin{equation}
\chi^{(2)}_m(2\omega; \omega, \omega) = -\text{i} a \frac{2 F \omega_0^4 A \mu_0 \omega^3}{D(\omega)^2 D(2\omega)},
\end{equation}
where $a = \pm 0.198$ V$^{-1}$ is the fitted strength of the nonlinearity, $A=113$ mm$^2$ is the ring's area, and $D(\omega)=\left(\omega_0^2 - \text{i}\gamma\omega - \omega^2\right)$ is the Lorentz denominator. The sign of $a$ is determined by the orientation of the varactor diode in the VLSRR, reflecting the fact that reversing the diode direction is analogous to reversing the orientation of the nonlinear metacrystal, and thus is expected, by symmetry arguments, to induce a $\pi$ shift in the phase of the even-order nonlinearities. Note that the magnitude of $a$ is close to the predicted value of $\frac{M}{2V_p}=0.267$ V$^{-1}$~\cite{Poutrina2010Analysis}.

The dispersion of the TE$_{10}$ mode of the metallic waveguide is modeled by
\begin{equation}
k_z(\omega) = \pm \omega \sqrt{ \mu_x \left(\epsilon_y  \left(1 - \frac{\omega_c^2}{\omega^2}\right) \right)}, \label{wavevector}
\end{equation}
where $k_z(\omega)$ is the longitudinal wavevector, $\omega_c = \frac{\pi}{b\sqrt{\epsilon_y \mu_0}}=2\pi\times674$ MHz is the cutoff frequency, and $b=15$ cm is the waveguide width. As in typical negative-index media, the negative sign is chosen when both terms in the square root are negative, i.e., when $\mu_x<0$ and $\epsilon_y  \left(1 - \frac{\omega_c^2}{\omega^2}\right)<0$. The former occurs just above the resonance frequency of the VLSRR, while the latter requires operation of the waveguide below cutoff, so that we can expect a backward-wave to exist for frequencies of roughly 615 MHz $< \omega/2\pi <$ 674 MHz. Furthermore, for a FF below cutoff, it can be shown that only the TE$_{10}$ mode will propagate at the SH frequency, simplifying our analysis. For coupling between the TE$_{10}$ modes at the FF and SH, we obtain an overlap integral of $\Gamma \approx 1.2$~\cite{Gunter2000Nonlinear}.

\begin{figure}[!htbp]
\center
\includegraphics[width=.4\textwidth]{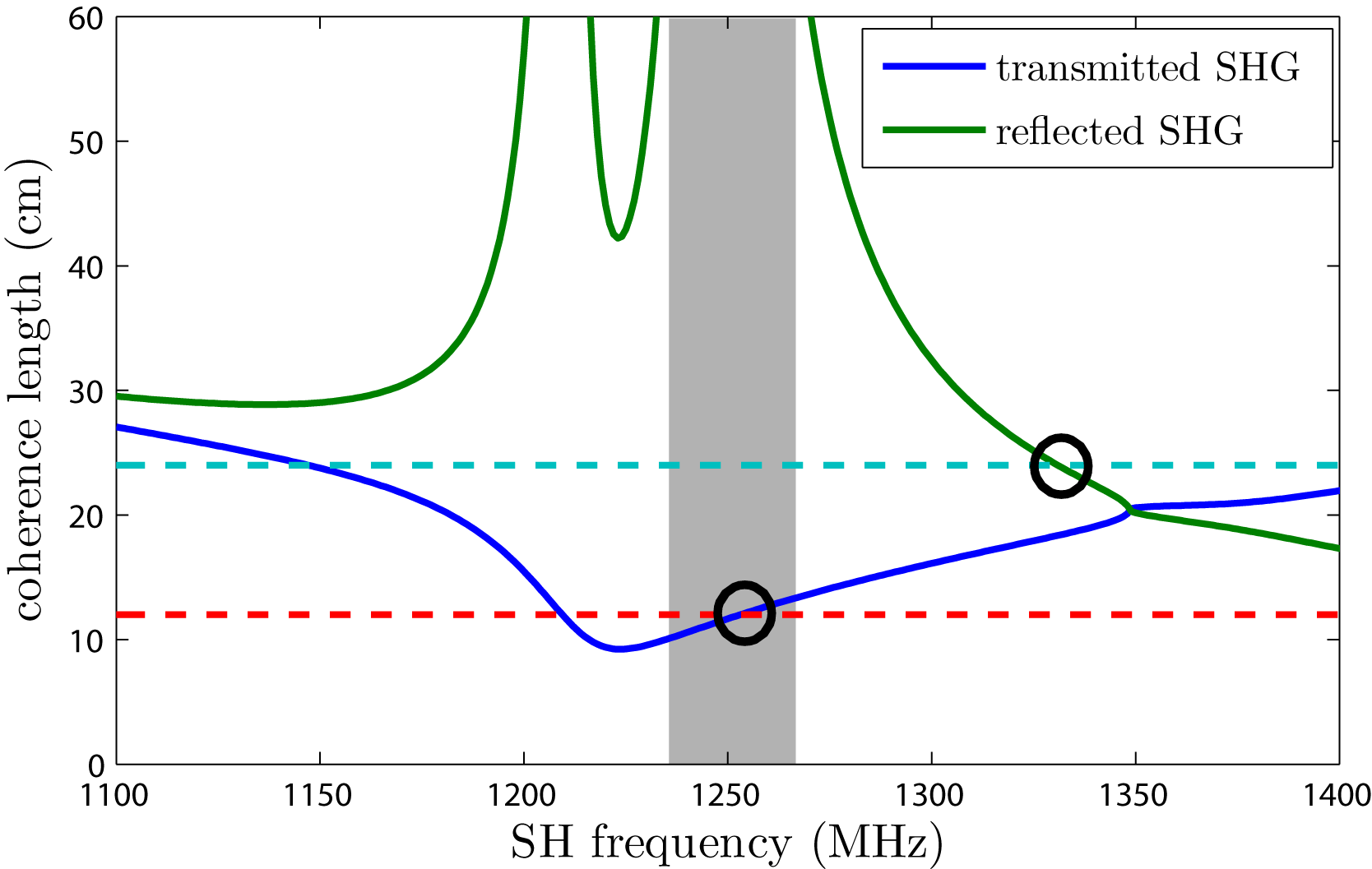}
\caption{(Color online) Plot of the calculated coherence lengths for transmitted (blue) and reflected (green) SHG in the VLSRR loaded waveguide. The dashed lines at 12 and 24 cm correspond to the two poling periods used in this letter, while the gray band indicates nonlinear-optical mirror behavior.}
\label{lcoh}
\end{figure}

The phase matching effects in our VLSRR loaded waveguide can be examined by employing the above experimentally fitted models to calculate the SHG phase mismatch and coherence lengths. For example, the phase mismatch for SHG in a homogeneous medium is given by
\begin{equation}
\Delta k = \pm k_z(2\omega) - 2k_z(\omega), \label{mismatch}
\end{equation}
where it is assumed that the energy of the FF wave propagates in the positive $z$ direction, and the $\pm$ refers to a positively (transmitted) or negatively (reflected) propagating SH wave. Thus, assuming the SH is far from resonance, it is clear that a forward-wave ($k_z(\omega)>0$) at the FF will tend to phase match with a transmitted SH wave, as is seen in conventional nonlinear optics. A backward-wave ($k_z(\omega)<0$) at the FF, on the other hand, will tend to phase match with a reflected SH wave, as predicted earlier~\cite{Agranovich2004Linear, Shadrivov2006Second-harmonic, Popov2006Negative-index}. This odd behavior can be seen more clearly by computing the coherence lengths, $L_\text{coh} = \frac{2\pi}{\Delta k}$, shown in Fig. \ref{lcoh} using Eqs. \eqref{wavevector} and \eqref{mismatch}. As expected, the coherence length for the transmitted SH is small (less than the interaction length), while the coherence length for the reflected SH is much larger throughout the negative-index band. The reflected phase-matched region, shown in gray in Fig. \ref{lcoh}, represents the frequency range in which our NLMM acts as the nonlinear-optical mirror of Refs. \cite{Popov2006Negative-index} and \cite{Shadrivov2006Second-harmonic}, with the vast majority of the SH power traveling opposite to the FF.

In an inhomogeneous medium with some spatial periodicity, however, the phase matching condition is rewritten to include the corresponding reciprocal lattice vector. In periodically poled QPM, for example, the sign of the nonlinear susceptibility is reversed in neighboring layers with a period of $\Lambda$. Quasiphase matching then requires
\begin{equation}
\pm k_z(2\omega)=2k_z(\omega) + 2 m \pi/\Lambda, \label{QPM}
\end{equation}
or, equivalently, $\Lambda = m L_\text{coh}$, where $m$ is any integer. As mentioned earlier, poling in our system can be achieved by simple reorientation of appropriate sections of VLSSRs, analogous to domain engineering. Thus, taking $m=1$ in Fig. \ref{lcoh}, the intersection of the dashed lines (representing poling periods of 12 and 24 cm) with the coherence length curves indicate the frequencies at which Eq. \eqref{QPM} is expected to be satisfied, resulting in enhanced SHG output for the respective configuration.

Furthermore, we see that something interesting happens as the FF approaches the cutoff frequency. Near this frequency, $k_z(\omega)$ vanishes. Thus, Eq. \eqref{QPM} can be satisfied for both the reflected and transmitted modes by the same poling period. This is indicated in Fig. \ref{lcoh} as the point where the reflected and transmitted modes cross, corresponding to where the FF passes through the cutoff frequency. Thus, for poling periods in the proximity of this coherence length, we expect to see QPM enhancement of \textsl{both} the reflected and transmitted SH waves.

\begin{figure}[!htbp]
\center
\includegraphics[width=.43\textwidth]{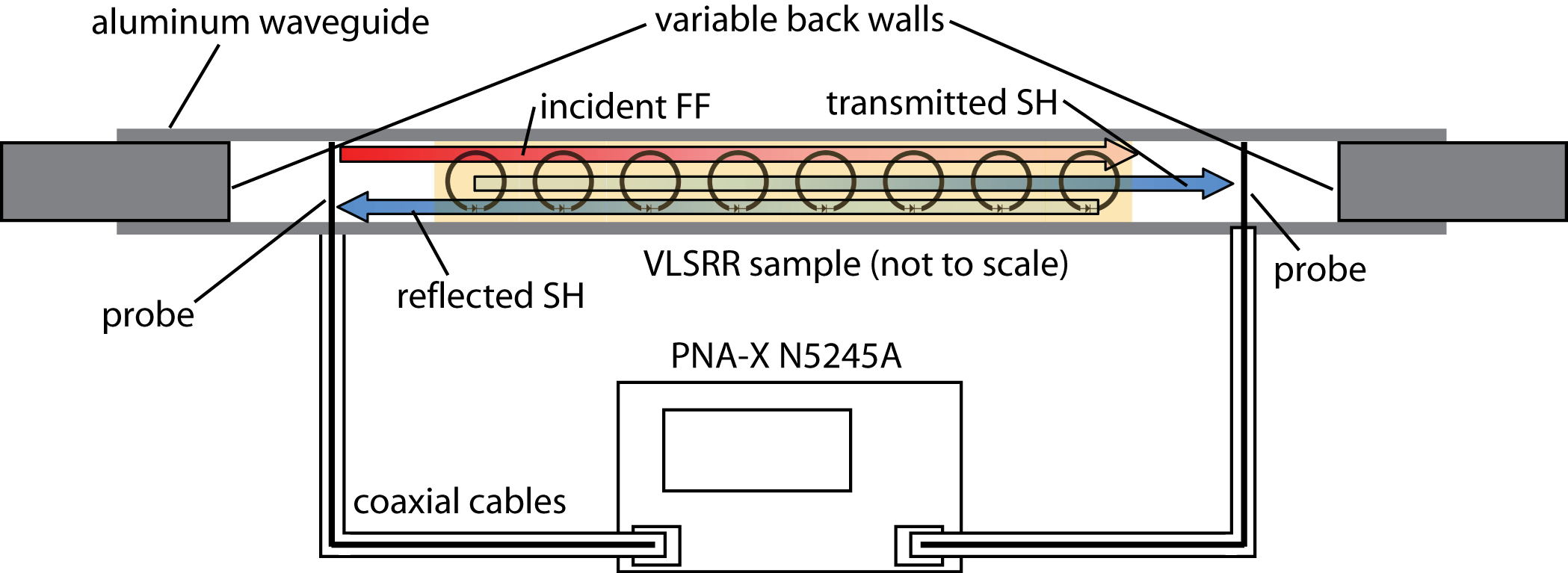}
\caption{(Color online) Diagram of the experimental setup employed to measure the SH spectrums, depicting a cross-section of the VLSRR loaded aluminum waveguide.}
\label{experiment}
\end{figure}

In order to inject and measure the FF and SH waves, respectively, the inner conductors (probes) of two coaxial cables are inserted into the waveguide on either side of the VLSRR sample. The probes are positioned 1.5 cm from the sample and covered with an alumina shell to reduce the impedance mismatch between the coaxial cable and the waveguide. Rectangular aluminum slabs with cross-section exactly equal to the waveguide are placed at a variable distance from the probes. By manually optimizing the distance between these back walls and the probes, we are able to reduce reflections at the coaxial-waveguide interface to less than 0.75 dB at the SH frequencies of interest, preserving the directionality of the generated SH wave. The FF, however, does not need to be matched because the much larger material losses in the FF range prevent significant back-reflection inside the sample. In other words, the vast majority of the incident FF power that is able to enter the sample travels in the forward direction and is absorbed by the NLMM before reaching the second interface, preserving the directionality of the FF wave. Thus, by pumping through one probe at the FF and measuring the SH power at both probes, we are able to simultaneously investigate the reflected and transmitted SHG in the VLSRR loaded waveguide. For clarity, a diagram of the experimental apparatus is shown in Fig. \ref{experiment}. An Agilent PNA-X N5245A network analyzer is used as both the source and receiver.

In order to validate our experimental results and interpretations, we employ nonlinear TMM calculations on a homogeneous slab with equivalent effective properties to the VLSRR loaded waveguide, sandwiched between semi-infinite slabs of vacuum, assuming the non-depleted pump approximation~\cite{Larouche2010retrieval, Rose2010Nonlinear}. The power incident on the sample at the FF is approximated by $P_i = P_0 - P_r$, where $P_r$ is the reflected power measured in the experiment and $P_0$ is the source's output power. The losses before and after the sample are assumed to be 0.75 dB at all frequencies. Although this system is clearly a rough approximation of the actual experimental setup, qualitative agreement is still expected, especially regarding the location and shape of the phase-matching enhancements.

\begin{figure}[!htbp]
\center
\includegraphics[width=.38\textwidth]{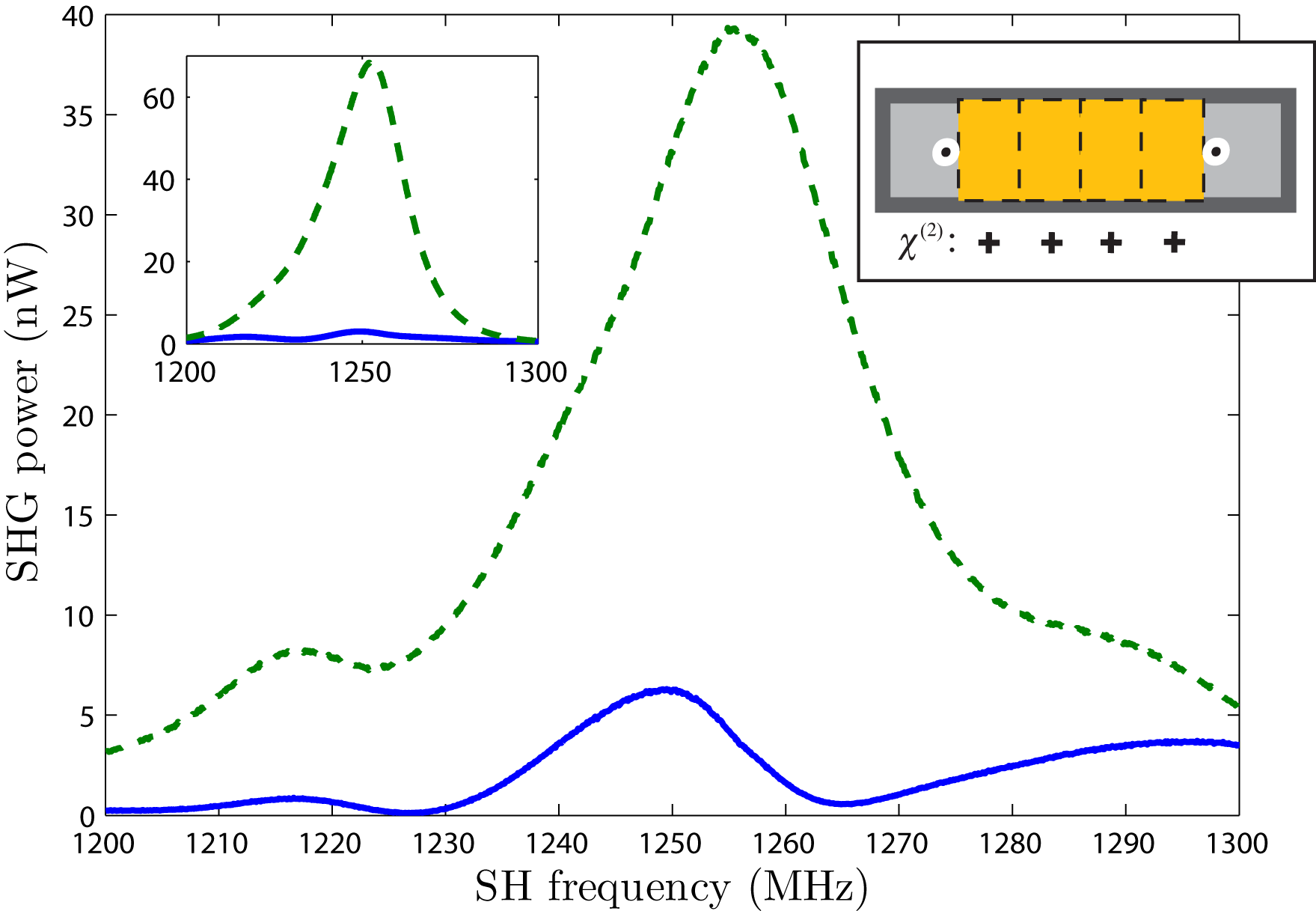}
\caption{(Color online) Comparison of the transmitted (solid blue) and reflected (dashed green) SHG powers when all sections are aligned. The right inset depicts the experimental configuration, and the left inset shows the TMM calculation.}
\label{reflect}
\end{figure}

First, we investigate phase-matching of the reflected SHG in a negative-index band. All four sections are aligned to simulate a homogeneous nonlinear medium. The source is swept from 600 to 650 MHz with an output power of -5 dBm. The resulting transmitted and reflected SH spectrums are plotted in Fig. \ref{reflect}. The plot shows significant enhancement in the SHG of the reflected wave over the transmitted, and is supported qualitatively by the inset showing the TMM result. Furthermore, the peak frequency of 1255 MHz lies well within the range predicted by the coherence length calculation of Fig. \ref{lcoh}. These results confirm the referenced theoretical studies, and in this configuration our NLMM can be considered the first realization of a nonlinear-optical mirror. Moreover, for the network analyzer's maximum output, we were able to display nonlinear-optical mirror action with a conversion efficiency as high as 1.5\%.

\begin{figure}[!htbp]
\center
\includegraphics[width=.38\textwidth]{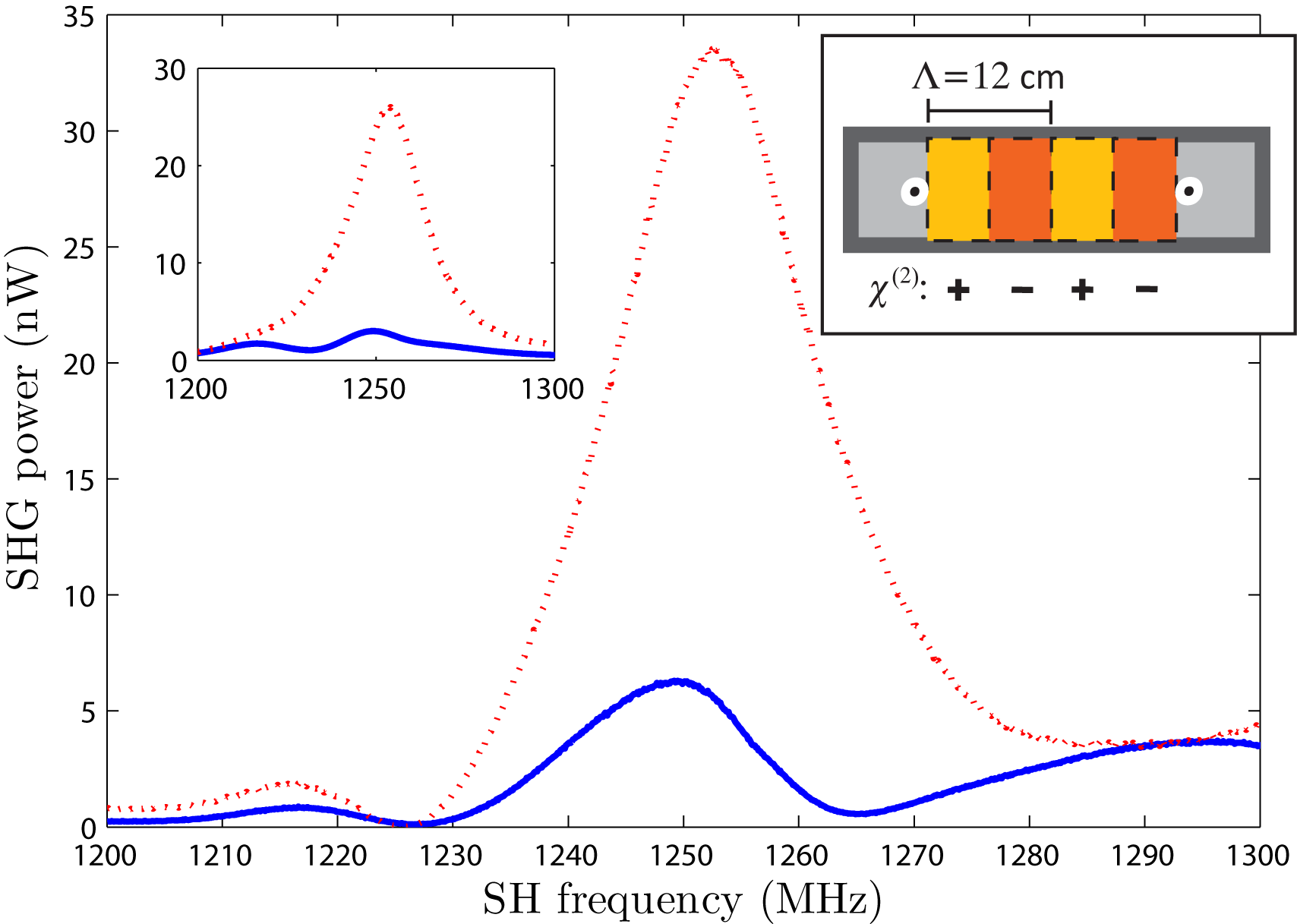}
\caption{(Color online) Comparison of the transmitted SHG power when the varactors are aligned (solid blue) and when they are periodically poled with $\Lambda=12$ cm (dotted red).}
\label{transmit}
\end{figure}

In the next experiment, the previous configuration is retained in all ways except that sections 2 and 4 are physically rotated by 180 degrees, effectively producing a periodically poled nonlinear crystal with $\Lambda=12$ cm. The transmitted SH spectrum of this QPM setup is compared to that of the original in Fig. \ref{transmit}. The transmitted SH power is greatly increased due to QPM, with the peak frequency of 1253 MHz in excellent agreement with Fig. \ref{lcoh}. Once again, qualitative agreement is found between the experimental data and the TMM calculations.

\begin{figure}[!htbp]
\center
\includegraphics[width=.38\textwidth]{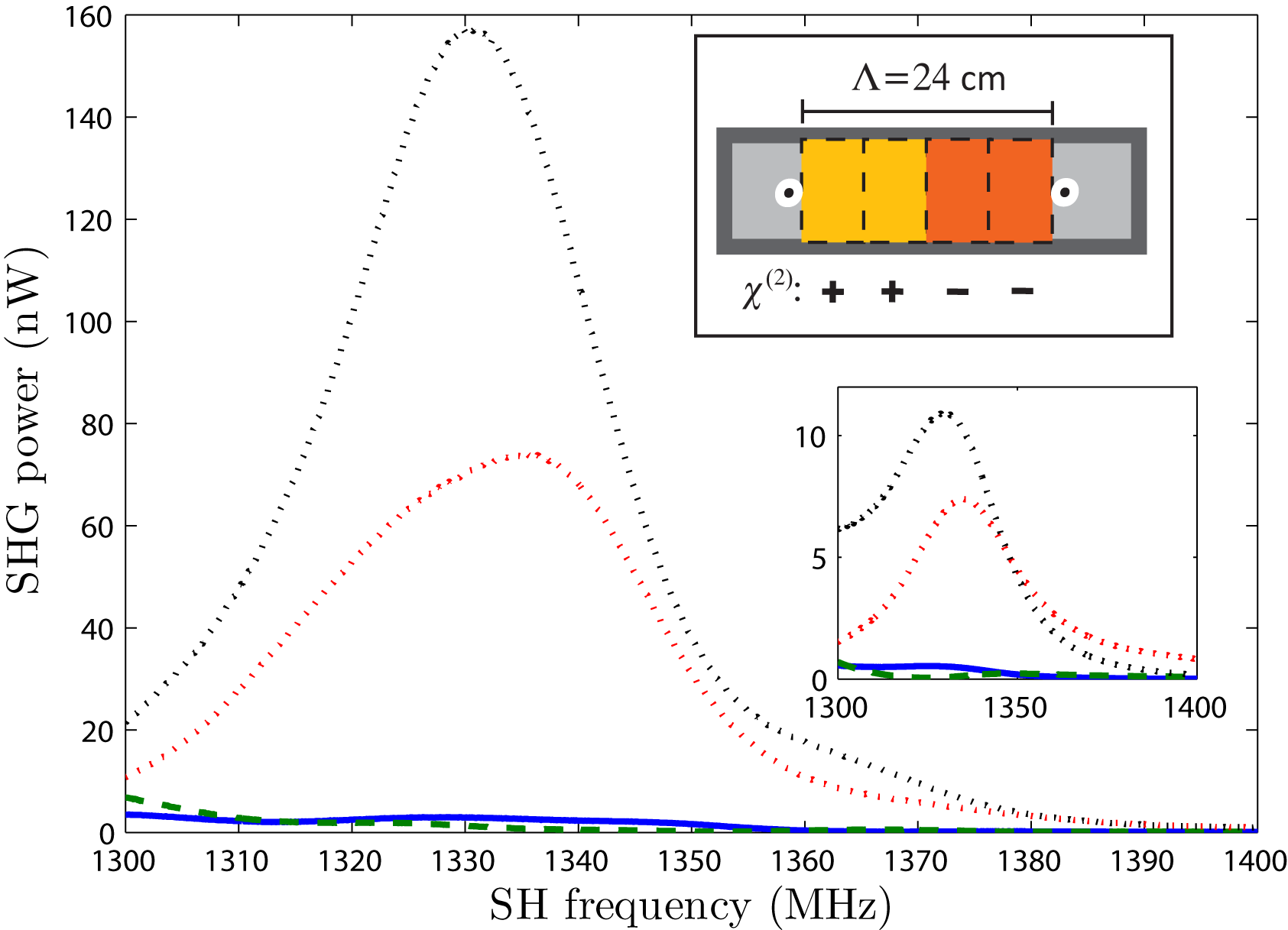}
\caption{(Color online) Comparison of the transmitted (solid blue and dotted red) and reflected (dashed green and dotted black) SHG power when the varactors are aligned and when they are periodically poled with $\Lambda=24$ cm, respectively.}
\label{both}
\end{figure}

In the final configuration, sections 1 and 2 are aligned and oriented opposite to sections 3 and 4, simulating a poling period of 24 cm. Since this poling period is close to the $k_z(\omega)=0$ crossing point in Fig. \ref{lcoh}, we can expect a strong enhancement of the reflected wave near a SH frequency of 1330 MHz, with a simultaneous (though weaker) enhancement of the transmitted wave. This is precisely what is shown in Fig. \ref{both}, with the SHG powers for the un-poled case plotted for comparison. While the inset again shows good qualitative agreement between the TMM and the experimental results, it should be noted that the peak emitted powers are off by an order of magnitude, whereas much better agreement is found in the previous two configurations. This is most likely due to the inherent sensitivity when approaching the near-zero frequency band, and it was confirmed in TMM calculations (not shown) that small changes in the linear properties produce large changes in the magnitude of the emitted radiation, but with similar qualitative behavior. Moreover, it has been shown that spatial dispersion, neglected here, can play an important role as the wavevector vanishes~\cite{Belov2003Strong}.

The exotic phase-matching effects shown here are made possible due to the unique linear properties accessible in NLMMs, in particular negative- and zero-index behavior. In addition, the use of varying periods to achieve periodic poling to switch between the various SHG schemes demonstrates the compatibility between NLMMs and QPM--a key result. We expect these results to be a necessary and vital step towards the realization of high efficiency nonlinear mirrors and other NLMM-based devices at microwave and terahertz frequencies.

This work was supported by the Air Force Office of Scientific Research (Contract No. FA9550-09-1-0562).

\bibliography{myrefs}
\end{document}